\documentclass[aps,prc,reprint,showpacs,superscriptaddress]{revtex4-1}

\usepackage{graphicx}
\usepackage{amsmath}

\begin{document}

\title{Using Light Charged Particles to Probe the Asymmetry Dependence of the Nuclear Caloric Curve}

\def \TAMUCI {Cyclotron Institute, Texas A\&M University, College Station, Texas, 77843, USA}
\def \TAMUChem {Chemistry Department, Texas A\&M University, College Station, Texas, 77843, USA}

\author{A.B. McIntosh}
\email[Corresponding Author: ]{amcintosh@comp.tamu.edu}
\affiliation{\TAMUCI}

\author{A. Bonasera}
\affiliation{\TAMUCI}
\affiliation{Laboratori Nazionali del Sud, INFN, I-95123 Catania, Italy}

\author{Z. Kohley}
\altaffiliation{Present Address: National Superconducting Cyclotron Laboratory, Michigan State University, East Lansing, Michigan 48824, USA}
\affiliation{\TAMUCI}
\affiliation{\TAMUChem}

\author{P.J. Cammarata}
\affiliation{\TAMUCI}
\affiliation{\TAMUChem}

\author{K. Hagel}
\affiliation{\TAMUCI}

\author{L. Heilborn}
\affiliation{\TAMUCI}
\affiliation{\TAMUChem}

\author{J. Mabiala}
\affiliation{\TAMUCI}

\author{L.W. May}
\affiliation{\TAMUCI}
\affiliation{\TAMUChem}

\author{P. Marini}
\altaffiliation{Present Address: GANIL, Bd Henri Becquerel, BP 55027 - 14076 CAEN Cedex 05, France}
\affiliation{\TAMUCI}

\author{A. Raphelt}
\affiliation{\TAMUCI}
\affiliation{\TAMUChem}

\author{G.A. Souliotis}
\affiliation{\TAMUCI}
\affiliation{Laboratory of Physical Chemistry, Department of Chemistry, National and Kapodistrian University of Athens, Athens GR-15771, Greece}

\author{S. Wuenschel}
\affiliation{\TAMUCI}
\affiliation{\TAMUChem}

\author{A. Zarrella}
\affiliation{\TAMUCI}
\affiliation{\TAMUChem}

\author{S.J. Yennello}
\affiliation{\TAMUCI}
\affiliation{\TAMUChem}

\date{\today}

\begin{abstract}
Recently, we observed a clear dependence of the nuclear caloric curve
on neutron-proton asymmetry $\frac{N-Z}{A}$
through examination of fully reconstructed equilibrated quasi-projectile sources
produced in heavy ion collisions at E/A = 35 MeV.
In the present work,
we extend our analysis
using multiple light charged particle probes of the temperature.
Temperatures are extracted with five distinct probes using a kinetic thermometer approach.
Additionally, temperatures are extracted using two probes
within a chemical thermometer approach (Albergo method).
All seven measurements show
a significant linear dependence of the source temperature on the source asymmetry.
For the kinetic thermometer,
the strength of the asymmetry dependence
varies with the probe particle species
in a way which is consistent with an average emission-time ordering.
\end{abstract}


\maketitle 


\section{Introduction}

The equation of state (EoS) of nuclei and nuclear matter
describes the emergent properties of a nuclear system,
i.e. the relation between the thermodynamic quantities,
which arise from the microscopic interactions between the constituent nucleons.
The nuclear EoS is of broad interest
for its importance in
nucleosynthesis, heavy ion collisions, supernovae dynamics, and neutron stars
\cite{Janka07,Danielewicz02,Li08,Lattimer04}.
To examine the EoS,
often the relation between only two thermodynamic quantities is examined.
These could be, for example, energy and density, pressure and temperature, density and temperature,
or temperature and energy; this last is commonly referred to as the caloric curve
and has been measured for many finite nuclear systems.
A prime example of this is shown in work by Pochodzalla et al.~\cite{Pochodzalla95}
where an observed plateau in the caloric curve was interpreted as a signature of a phase transition.
More recently, systematic analysis has shown that the caloric curve
depends on the mass $A=N+Z$ of the finite system \cite{Natowitz02}.

Currently, the largest uncertainty in the nuclear EoS
is the asymmetry energy (also referred to as the symmetry energy).
Although significant ongoing effort has been put forth
to constrain this,
very little of it has focused on the influence of the neutron-proton asymmetry,
$\frac{N-Z}{A}$,
on the nuclear caloric curve.

Our recent measurement \cite{McIntosh13}
showed a clear and strong dependence of nuclear temperatures on the asymmetry
using protons as probes.
It was shown
that an accurate reconstruction of the fragmenting quasi-projectile source
was essential to extract this dependence.
This paper extends our previous analysis using charged clusters as probes of the temperature.
All of these temperature probes, some kinetic and some chemical,
exhibit a significant dependence of the temperature on asymmetry.

Theoretical calculations provide conflicting predictions
of the dependence of the caloric curve on the asymmetry.
Some theoretical approaches
(the Thomas-Fermi model \cite{Kolomietz01} and the mononucleus model \cite{Hoel07})
predict that critical temperatures or limiting temperatures
should be higher for neutron-poor systems.
Other theoretical approaches
(the hot liquid drop model \cite{Besprosvany89},
the statistical multifragmentation model \cite{Ogul02},
and isospin-dependent quantum molecular dynamics \cite{Su11})
predict higher temperatures for neutron-rich systems.
The treatment of a ``gas'' phase in a model
is expected to impact the asymmetry dependence
of the temperature of the bulk system \cite{Hoel07,Besprosvany89},
and thus an observation of an asymmetry dependence
may support the physical picture of a nuclear liquid interacting with its vapor \cite{Hoel07}.
Moreover, observation of an asymmetry dependent temperature
may allow insight into the driving force of nuclear disassembly \cite{Sfienti09}.

Experimental efforts
to probe the asymmetry energy
in the nuclear equation of state,
such as Refs.~\cite{Colonna06,Shetty07,Marini12},
use isotopic fragment yields as an observable,
and make the assumption that the temperature is independent of the asymmetry.
Characterization of the asymmetry-dependence of the caloric curve
would allow a refined interpretation of this fragment yield data
(e.g. in the statistical interpretation of isoscaling).
Moreover, characterization of this asymmetry dependence may offer a new opportunity
to probe the asymmetry energy.

With one exception \cite{McIntosh13},
each previous analysis of experimental data to investigate this phenomena
has demonstrated either a small asymmetry dependence \cite{Sfienti09,Trautmann08}
or no discernible asymmetry dependence \cite{Wuenschel10,Wuenschel09_Thesis}.
However, an accurate selection of the asymmetry of the excited source
is crucial to discerning the temperature dependence \cite{McIntosh13,Wuenschel09_Thesis}.
In the present work,
we build on our recent observation,
now demonstrating for a variety of temperature probes
that nuclear temperatures exhibit a significant asymmetry dependence.

\section{Experiment, Event Selection and Reconstruction}

To study the dependence of the nuclear caloric curve on neutron-proton asymmetry,
collisions of $^{70}$Zn+$^{70}$Zn,
$^{64}$Zn+$^{64}$Zn, and
$^{64}$Ni+$^{64}$Ni
at E/A = 35 MeV
were studied.
For details of the experiment, see Refs.~\cite{Kohley10,Kohley11_LCP}.
Charged particles and free neutrons produced in the reactions
were measured in the
NIMROD-ISiS $4\pi$ detector array \cite{Wuenschel09_Thesis,Wuenschel09_NIM}.
The energy resolution achieved allowed excellent isotopic resolution of
charged particles up to Z=17.
For events in which all charged particles are isotopically identified,
the quasi-projectile
(QP, the primary excited fragment
that exists momentarily after a non-central collision)
was reconstructed
using the charged particles and free neutrons.
Thus, the reconstruction includes determination of the QP composition, both A and Z.

To examine the caloric curve,
it is desired to select equilibrated quasi-projectile sources.
To this end, three cuts are made
on the particle and event characteristics.
Similar cuts, which have been used previously
\cite{McIntosh13,Marini12,Wuenschel10,Wuenschel09_Thesis,Wuenschel09_Isoscaling,Steckmeyer01},
have been shown to select events
with properties consistent with equilibrated QP sources.
To exclude fragments that clearly do not originate from a QP,
the fragment velocity in the beam direction $v_z$
is restricted
relative to the velocity of the heavy residue $v_{z,res}$.
The heavy residue is defined as the largest fragment measured in the event,
and thus is the largest decay product of the QP.
The accepted window on $\frac{v_z}{v_{z,res}}$ is $1 \pm 0.65$ for Z=1,
$1 \pm 0.60$ for Z=2,
and $1 \pm 0.45$ for Z$\geq$3.
This eliminates fast pre-equilibrium fragments, fragments originating
from the quasi-target (QT), and some fragments which may be produced
non-statistically between the QP and QT.
Selection of a narrow mass range is important due to the mass-dependence of the caloric curve,
and because a large range of source masses could introduce fluctuations which impact the
momentum quadrupole fluctuation thermometer.
One would like to select a QP mass range not too far from the projectile mass,
while still selecting an ensemble of events with sufficient statistics for the analysis.
The mass of the reconstructed QP is required to be $48 \leq A \leq 52$,
satisfying the requirements.
The selection of thermally equilibrated QPs is achieved through the selection of spherical events,
since the shape degree of freedom is quite slow to equilibrate
compared to the thermal degree of freedom \cite{Hinde89a,Hinde89b}.
Spherical QPs are selected by a constraint on the
longitudinal momentum $p_z$
and transverse momentum $p_t$
of the fragments comprising the QP:
$-0.3 \leq log_{10}(Q_{shape}) \leq 0.3$
where $Q_{shape} = \frac{\sum{p_z^2}}{\sum{p_t^2}}$
with the sums extending over all fragments of the QP.
In this way, we have obtained a cohort of quasi-projectiles,
tightly selected on mass,
with properties consistent with thermal equilibration,
and with known neutron-proton asymmetry.

The uncertainty in the QP composition arises mainly from the free neutron measurement.
We employ a model which deduces the number of free neutrons emitted by the QP
from the measured number of total neutrons, background, and the efficiencies for measuring
neutrons produced from QP and QT sources \cite{Wuenschel09_Isoscaling,Wuenschel10,Wuenschel09_Thesis}.
We have investigated the efficiency of the neutron ball detector
through detailed simulations \cite{Marini12b,Wada04}.
The uncertainty in the free neutron measurement has two effects:
the assignment of a QP to an incorrect bin in neutron-proton asymmetry,
and an error in the calculation of the excitation energy.
The uncertainty in the free neutron multiplicity
is due primarily to the efficiency of the neutron detector ($\approx$70\%),
and to a lesser extent the yield of background signals in the neutron detector.
The efficiency causes an increase in the width of the measured neutron distribution.
This width of the neutron distribution given a fixed number of initial neutrons
depends on the initial number of neutrons
but is less than 2.1 for the multiplicities encountered in this analysis \cite{Marini12b}.
Subtracting this in quadrature from the width of the raw distribution (5.36)
gives a ``true'' width of 4.97.
Thus the increase in the width of the neutron distribution
due to the efficiency is at most 9\%
at the highest excitation energies.
The measured background multiplicity distribution has a width
that is narrower than the raw distribution by a factor of more than three.
Subtracting these widths in quadrature reveals that
the increase in the width due to the background is less than 6\%.
The combined effect of the efficiency and the background (added in quadrature)
is thus on the order of 11\%.
Given this uncertainty, assignment of a QP to an asymmetry bin is rather precise.

The excitation energy of the QP was deduced using
the measured free neutron multiplicity,
the charged particle kinetic energies in the transverse direction,
and the Q-value of the breakup.
Excitation energies above $E^*/A$ = 2 MeV
are well measured with the present detector array.
The average kinetic energy per free neutron ascribed to the QP
is given by $2.2 + 1.25 (E_{CP}/A_{CP})$
where $E_{CP}$ is the total energy of the charged particles in the transverse direction
and $A_{CP}$ is the mass of the QP calculated using only charged particles.
This relation is determined from the experimental
Coulomb-shifted transverse kinetic energy spectra of protons.
The aptness of the Coulomb shift
was seen by applying the shift to the
triton ($^3$H) and helion ($^3$He) transverse kinetic energy spectra,
which completely overlap when the shift is applied.
For an excitation $E_{CP}/A_{CP}$ of 6 MeV,
the average neutron kinetic energy ascribed to the QP is 9.7 MeV
and thus, for a QP with 50 nucleons where 5 become free neutrons,
the free neutrons contribute 0.97 MeV per nucleon to the total excitation energy per nucleon.
An error of 11\% on the free neutron multiplicity corresponds to an error of
0.11 MeV per nucleon.
This error is significantly less than the spacing between
even the closest caloric curves selected on asymmetry
(see section \ref{sec:Results}).
We considered the possibility of an auto-correlation between the
portion of the excitation energy due to the neutrons and the observed dependence of the caloric curve.
For any physically realistic values of neutron kinetic energy and multiplicity,
the asymmetry dependence of the caloric curve is robust.
In summary, the uncertainties in the QP composition and in the QP excitation energy
due to the free neutron measurement
are sufficiently small
that they do not significantly bias the results reported in this paper.

The temperatures of the QPs are calculated using two different methods.
The first method is the momentum quadrupole fluctuation (MQF) thermometer \cite{Zheng11_PLB},
which has been previously used to examine
temperatures of nuclei \cite{Wuenschel10,Wuenschel09_Thesis,Stein12x,McIntosh13,Mabiala12}.
The momentum quadrupole is defined as
\begin{equation}
Q_{xy} = p_{x}^{2} - p_{y}^{2},
\end{equation}
using the transverse components $p_x$ and $p_y$ of the particle's momentum
in the frame of the QP source.
If the approximation is made that
the emitted (probe) particles are described by a Maxwell-Boltzmann distribution,
the variance of $Q_{xy}$ is related to the temperature by
\begin{equation}
\langle \sigma_{xy}^2 \rangle = 4m^2T^2,
\end{equation}
where $m$ is the probe particle mass \cite{Wuenschel10, Zheng11_PLB}.
Since the longitudinal component, $p_{z}$,
may be impacted by the velocity cut which was imposed to exclude fragments which do not
originate from an equilibrated QP source,
only the transverse components are used to calculate the quadrupole.
For this analysis,
protons, deuterons, tritons, helions ($^3$He), and alpha particles
are used separately as probes
(denoted p, d, t, h, and $\alpha$ respectively).
Since the thermal energy in the primary clusters
is of course significantly less than that in the QP,
the width of the momentum quadrupole distribution
is dominated by the QP breakup.
Therefore it is expected that the effects of secondary decay
on this thermometer are small.

Temperatures of the excited QPs are also calculated
using yield ratios
following the method of Albergo \emph{et al.} \cite{Albergo85}.
Here, the temperature is calculated as
\begin{subequations}
\begin{align}
T_{raw} = \frac{B}{\ln\left(aR\right)}, \\
 T = \frac{1}{\frac{1}{T_{raw}}-\frac{\ln\kappa}{B}},
\end{align}
\end{subequations}
where $R$ is a double ratio of isotopic yields,
$B$ is a difference of binding energies,
$a$ is a mass-weighted spin degeneracy factor,
and $\frac{\ln\kappa}{B}$ is a correction accounting for secondary decay
(for details, see, for example, Ref.~\cite{Xi99}).
Two combinations of isotopic yields were used to extract temperatures.
For $R=\frac{Y(d)/Y(t)}{Y(h)/Y(\alpha)}$,
the constants are $B=14.32$ MeV, $a=1.59$, and $\frac{\ln\kappa}{B}=0.0097$ MeV$^{-1}$;
for $R=\frac{Y(^6Li)/Y(^7Li)}{Y(h)/Y(\alpha)}$,
the constants are
$B=13.32$ MeV, $a=2.18$, and $\frac{\ln\kappa}{B}=-0.0051$ MeV$^{-1}$.

The two thermometers to be employed here operate in fundamentally different ways.
The MQF thermometer is a kinetic thermometer,
using the momentum of emitted particles to determine the temperature.
The Albergo thermometer is a chemical thermometer which uses the fragment yields.
As such, they measure different apparent temperatures.
Differences in the apparent temperature
as measured by different thermometers has been a focus of study for many years.
It is understood that certain kinetic thermometers may differ from
chemical thermometers due to a variety of factors
such as the Fermi motion of nucleons
(see e.g. Refs.~\cite{Bauer95,Kelic06}).

It is not the purpose of this paper to explore differences between thermometers.
Our purpose is to provide experimental evidence that regardless of the thermometer employed
or the the value of the temperature extracted,
the extracted temperature depends on the composition of the source.

\section{Results and Discussion}
\label{sec:Results}

\begin{figure}
\centering
\includegraphics{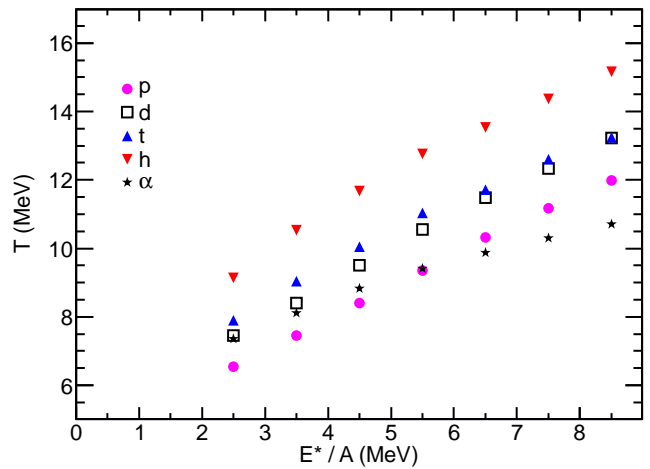} 
\caption{\label{fig:CompareLCPTemperature}
(Color online)
Momentum quadrupole fluctuation temperatures for
sources with mass $48 \leq A \leq 52$ and all asymmetries,
extracted with light charged particle probes.
}
\end{figure}

\begin{figure*}
\centering
\includegraphics[scale=1.04]{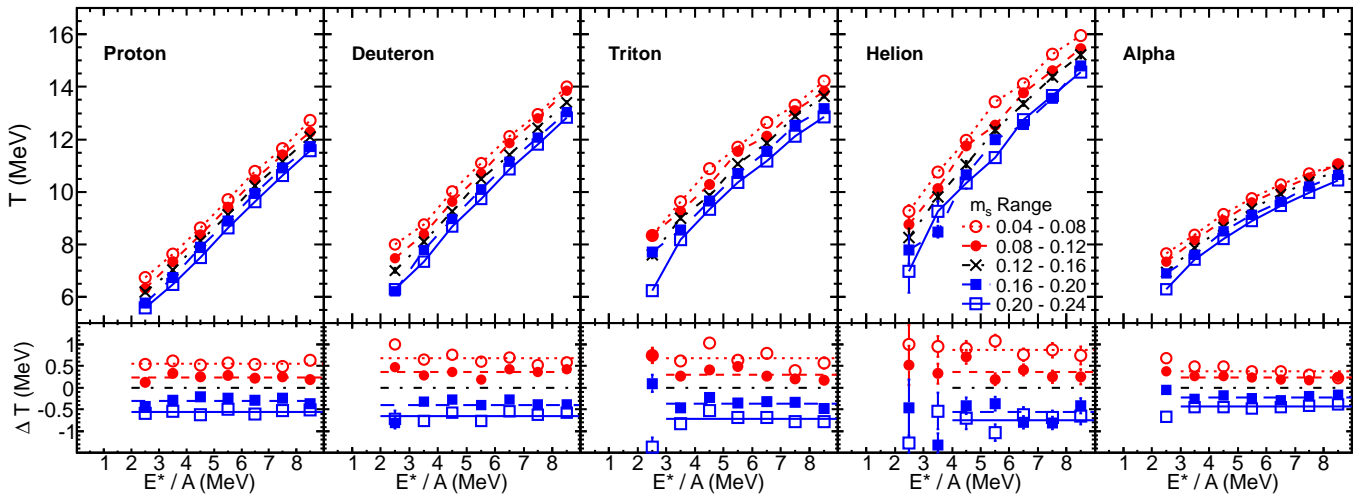} 
\caption{\label{fig:TClas_SelectMsMQFT_5x2}
(Color online)
Top row: caloric curves for light charged particles,
selected on source asymmetry ($m_s = \frac{N-Z}{A}$).
Bottom row: temperature difference
between each caloric curve and the middle caloric curve ($0.12 < m_s \leq 0.16$)
which is used as a reference.
The horizontal lines correspond to the average difference
over the indicated range in excitation.
Error bars
represent the statistical uncertainties.
}
\end{figure*}

Figure~\ref{fig:CompareLCPTemperature} shows the
temperature of the QP,
as a function of the excitation energy per nucleon ($E^*/A$) of the QP,
measured with the MQF thermometer
using protons, deuterons, tritons, helions and $\alpha$ particles as probe particles.
Temperatures are calculated for 1 MeV-wide bins in excitation energy per nucleon.
The statistical errors are smaller than the size of the points.
For all particle types examined,
the temperature shows a monotonic increase with $E^*/A$.
At $E^*/A = 2.5$ MeV,
the temperatures are between $5.5$ and $9.5$ MeV;
by $E^*/A = 8.5$ MeV,
the temperatures have risen to between $10.5$ and $15.5$ MeV.
The temperatures for the different light charged particles differ
by up to 5 MeV.

There is an overall ordering to these caloric curves.
For a given excitation energy per nucleon above 5.5 MeV
(where the bulk of the yield is),
the $\alpha$ particles show the lowest temperature,
followed by the protons,
deuterons, tritons, and, lastly, helions.
Below 5.5 MeV, the order is the same
except that the $\alpha$ particle and proton temperatures are inverted.
The ordering of the temperatures for the different LCP species are summarized by the relation
\begin{equation}
T_{\alpha}, T_{p} < T_{d} < T_{t} < T_{h}.
\end{equation}
This ordering is consistent with a previous analysis \cite{Wuenschel10},
which used these same LCP probes to measure the temperature
with the MQF method
for reconstructed sources produced in reactions of Kr+Ni at E/A = 35 MeV.

The ordering observed in Fig.~\ref{fig:CompareLCPTemperature}
may be due to different average emission times
for the different particles \cite{Hudan03,Chen03,Ghetti06,Kohley12}.
The tritons and helions are ``expensive'' particles,
as the energy cost (Q-value) for their emission from an excited source
($Q \approx 20$ MeV)
is high relative to protons and $\alpha$ particles
($Q \approx 10$ MeV).
Expensive particles are most likely to be emitted in the early stages
of the de-excitation of the source
when the available excitation energy is high.
Therefore, the temperature exhibited by the expensive particles
reflects the temperature of the source only
at the early stages of the de-excitation,
when the source is hottest.
Less expensive particles can be emitted
throughout the de-excitation.
Hence, the average temperature
probed by the less expensive particles is lower.
This emission-time-order scenario
is consistent with previous work \cite{Kohley12},
which used scaled transverse LCP flow
in reactions of Zn+Zn and Ni+Ni reactions at E/A = 35 MeV
in conjunction with
a molecular dynamics simulation (CoMD)
to extract the average emission order of the LCPs.
If different particles probe different space-time regions,
then Coulomb effects might slightly influence the results as well.
The persistence of the ordering at the highest excitation energies
(where very rapid breakup of the source may occur)
suggests that the time-ordering may not be a complete explanation.
Alternately, the temperature ordering
may be indicative of differences in the average density
of the source from which the probe particles originate.
Such density fluctuations are not at odds with the selection of an equilibrated source.
In fact, fragment production would not be possible in the absence of such fluctuations,
since fragments are essentially a high density region
in a low-density or zero-density background.

A plateau in the caloric curve has been observed previously,
and interpreted as a signature of a phase transition \cite{Pochodzalla95,Kelic06,Natowitz02}.
The lack of a plateau in the present data \cite{McIntosh13}, however,
is not unexpected.
A plateau is not as well defined for such small sources ($A \approx 50$)
as it is for heavier sources \cite{Natowitz02}.
Since the density is not constrained here,
a varying density may mask the plateau \cite{Borderie12,Mabiala12}.
Finally,
the plateau might lie above
the excitation energies measured in this experiment \cite{Natowitz02}.

Figure~\ref{fig:TClas_SelectMsMQFT_5x2} (top row, second panel)
shows the temperature as a function of excitation energy per nucleon ($E^*/A$)
of the QP as determined with the MQF thermometer
using deuterons as the probe particle.
Data points are plotted for 1 MeV-wide bins in excitation energy per nucleon.
For clarity, the points are connected with lines to guide the eye.
The error bars correspond to the statistical uncertainty
and, where not visible, are smaller than the points.
The temperature shows a monotonic increase with excitation energy.
At $E^*/A$ = 2.5 MeV,
the temperatures are around 7 MeV;
by $E^*/A$ = 8.5 MeV,
the temperatures have risen to around 13 MeV.
Each curve corresponds to a narrow selection in
the asymmetry of the source, $m_s = \frac{N_s-Z_s}{A_s}$,
where $N_s$, $Z_s$, and $A_s$ are the neutron number,
proton number, and mass number, respectively, of the source.
The $m_s$ distribution is peaked around 0.15
and is quite broad, having a standard deviation around 0.07.
The distribution is divided into bins of equal width
as indicated in the legend.
The average asymmetries for the selections are
0.0640, 0.0988, 0.1370, 0.1758, and 0.2145.
The caloric curve is observed to depend on
the asymmetry of the source.
Increasing the neutron content of the QP source
shifts the caloric curve to lower temperatures.
In fact, the caloric curves for the five source asymmetries
are approximately parallel and equally spaced.
An increase of 0.15 units in $m_s$
corresponds to a decrease in the temperature
by about 1.3 MeV.
The top left panel of Fig.~\ref{fig:TClas_SelectMsMQFT_5x2}
shows the caloric curves extracted with the MQF thermometer
using protons as the probe particles.
This was show in Ref.~\cite{McIntosh13}, and is shown here for completeness.
The three remaining panels of the top row of Fig.~\ref{fig:TClas_SelectMsMQFT_5x2}
show the caloric curves extracted with the MQF thermometer
using tritons, helions and $\alpha$ particles.
Again, each curve corresponds to a narrow selection in $m_s$.
For all LCP probes,
the caloric curve shifts to lower temperature for increasing neutron content.
For any given probe particle,
the curves are approximately parallel and equally spaced
within statistical uncertainties.

The impact of the QP mass selection ($48 \leq A_{QP} \leq 52$)
has been investigated.
It was considered that a range of masses
could ``artificially'' broaden the momentum quadrupole distributions,
making the extracted temperatures appear higher than they actually are.
It was observed that the selection of a single mass ($A=50$)
results in the same temperatures as shown here,
though the statistical uncertainties are of course larger.
Therefore, the width of the QP mass selection used in this analysis
does not impact the extracted temperatures or their dependence on $m_s$.

\begin{figure}
\centering
\includegraphics[scale=0.96]{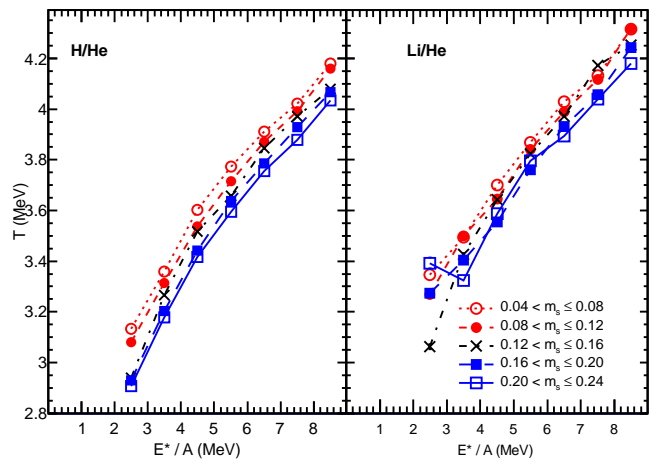} 
\caption{\label{fig:Albergo_tdha_HeLi}
(Color online)
Albergo yield ratio temperatures 
extracted with deuterons, tritons, helions and $\alpha$ particles (left panel),
and extracted with $^6$Li, $^7$Li, helions and $\alpha$ particles (right panel).
The different curves correspond to narrow selections on the source asymmetry.
}
\end{figure}

The shift in temperature due to the asymmetry of the source
is examined more closely in the lower row of Fig.~\ref{fig:TClas_SelectMsMQFT_5x2}.
The central $m_s$ bin ($0.12 < m_s \leq 0.16$) has been used as a reference.
The difference in temperature between each caloric curve and the reference
is plotted as a function of the excitation energy per nucleon.
For all LCP probes,
the differences are relatively constant with excitation energy.
The average $\Delta T$ for each pair of $m_s$ bins is indicated by the horizontal lines;
the range used for averaging is indicated by the endpoints of the lines.
Though the statistical fluctuations
vary for the different LCP probes,
the same trends are observed
in all cases.
For the $\alpha$ particles,
there may be a spreading of the caloric curves
at low excitation energy;
however, above $E^*/A=3$ MeV, the difference is approximately constant.
The constancy shows that the shift in the caloric curves
with asymmetry is independent of excitation energy.

We have previously shown the importance of selecting the asymmetry of the QP
rather than the asymmetry of the initial system \cite{McIntosh13}.
The temperature only slightly changes as the
asymmetry of the initial system is varied,
but the temperature does show a significant variation
with the asymmetry of the QP.
The method of asymmetry selection may account for other previous investigation not observing
an asymmetry-dependent temperature \cite{Wuenschel10,Wuenschel09_Thesis}.
As the QP de-excites, each particle emission changes its composition slightly.
If it were possible, one would correlate the temperature probed by the emitted particles
to the composition of the QP at the time the particles were produced.
Since this is not possible experimentally, our method determines the initial composition of the QP,
and correlates this with the temperature probed by the emitted particles,
which are emitted over a range of times.
Using the measured initial composition instead of the composition at the instant of particle emission
could mask the asymmetry-temperature correlation somewhat.
Therefore, the true dependence of the nuclear temperature on the asymmetry may be even greater than
the considerable dependence observed here.

Figure~\ref{fig:Albergo_tdha_HeLi}
shows caloric curves obtained with the Albergo thermometer
using d, t, h, $\alpha$ yields (left panel)
and using $^6$Li, $^7$Li, h, $\alpha$ (right panel).
These temperatures are lower than those obtained with the MQF method,
rising from around $T = 3$ MeV at $ E^*/A = 2$ MeV to
around $T = 4$ MeV at $E^*/A = 8$ MeV.
The difference between these temperatures
and those extracted with the MQF method
is a result of the different nature of the two thermometers (chemical vs kinetic)
as previously discussed.
The many methods of extracting nuclear temperatures
are known to yield disparate values for a variety of reasons
(see e.g. Ref.~\cite{Kelic06}).
The question at hand is whether, for a given thermometer,
the temperature can be observed to depend on the asymmetry.
The different curves in Fig.~\ref{fig:Albergo_tdha_HeLi}
correspond to selections on the asymmetry
of the isotopically reconstructed QP.
The temperatures extracted for neutron poor sources
are consistently higher than for neutron rich sources.
An increase of 0.15 units in $m_{s}$
corresponds to a decrease in the Albergo temperature
on the order of 0.15 MeV - small,
but statistically significant.
This trend is consistent with the results
of the MQF method.

\begin{figure}
\centering
\includegraphics{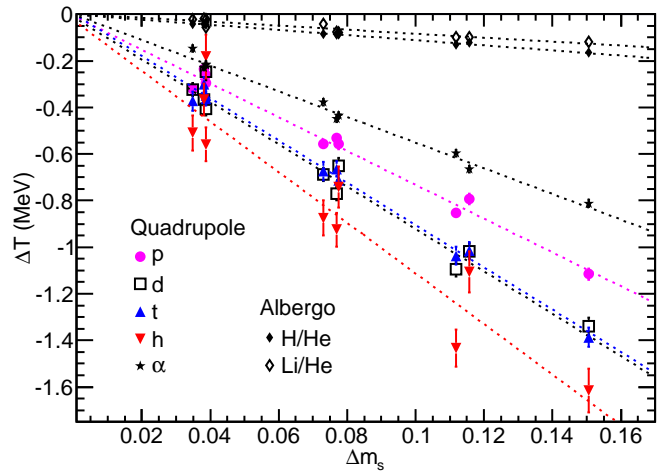} 
\caption{\label{fig:TempDiff_DeltaTvsMs_Lots}
(Color online)
Change in temperature $\Delta T$ with changing asymmetry $m_s$
for the five different probes of the momentum quadrupole fluctuation temperatures
and two methods of extracting the Albergo yield ratio temperature.
}
\end{figure}

To examine in more detail the temperature shift of the caloric curve
with changing source asymmetry,
we plot $\Delta T$ vs $\Delta m_s$
in Fig.~\ref{fig:TempDiff_DeltaTvsMs_Lots}.
This is obtained in the following way.
For each possible pairing (10 pairings total) of the five caloric curves (one for each $m_s$ bin),
the temperature difference as a function of excitation energy is calculated.
These differences are approximately constant as a function of excitation energy
as seen in the lower row of Fig.~\ref{fig:TClas_SelectMsMQFT_5x2};
the excitation-averaged $\Delta T$ is plotted in Fig.~\ref{fig:TempDiff_DeltaTvsMs_Lots}.
For each possible pairing of the five caloric curves,
$\Delta m_s$ corresponds to the difference in the mean asymmetries
of the pair of curves.
The results obtained with the MQF thermometer
using protons, deuterons, tritons, helions, and $\alpha$ particles are
shown with the circles, squares, up-triangles, down-triangles, and stars respectively;
the results from the two Albergo thermometers are shown with the open and closed diamonds.
The strength of the correlation for the Albergo method is weaker
than it is for the MQF method.
This may be a result of the smaller value of the temperature
for the Albergo thermometer as compared to the MQF thermometer,
and may as well be due to the different nature of the thermometers (chemical versus kinetic).
For every method and probe, the negative correlation follows a linear trend.
The slopes of linear fits to the data are
-7.3 $\pm$ 0.2, -9.2 $\pm$ 0.3, -9.4 $\pm$ 0.3, -10.6 $\pm$ 0.7, and -5.5 $\pm$ 0.1
MeV per unit change in $m_s$
for the p, d, t, h, and $\alpha$ particle probes.
The slopes of the linear fits are -1.01 $\pm$ 0.07 and -0.94 $\pm$ 0.15
MeV per unit change in $m_s$
for the H/He and Li/He Albergo probes respectively.
These linear fits describe the data well over a broad range in source asymmetry.

The strength of the $\Delta T / \Delta m_s$ correlation
for the MQF thermometer
shown in Fig.~\ref{fig:TempDiff_DeltaTvsMs_Lots}
depends on the probe particle.
Interestingly, the slopes are ordered in the same way by particle
that the absolute values of the temperatures are ordered
for $E^*/A > 5.5$ MeV (Fig.~\ref{fig:CompareLCPTemperature}).
This region in excitation is where the bulk of the yield is.
The helions, which exhibit the highest temperatures,
show the strongest dependence on $\Delta m_s$;
the protons and $\alpha$ particles,
with the lowest temperatures,
show the weakest dependence on $\Delta m_s$.
The ordering of the slopes
is summarized by the relation
\begin{equation}
t_{\alpha} < t_{p} < t_{d} < t_{t} < t_{h}.
\end{equation}
The reason for the ordering of the slopes is not yet established.
It is possible that the ordering is a direct result of the emission order.
In this scenario, the particles emitted earliest (helions, tritons)
would be coming from a source whose composition is experimentally well defined.
Particles which are on average emitted later (alpha, protons)
would be emerging from a source whose composition has changed somewhat
due to preceding particle emissions.
Thus, the correlation between
the difference in the extracted QP temperature ($\Delta T$)
for a given difference in initial QP asymmetry ($\Delta m_s$)
is least obscured for particles emitted at the earliest times.

The slope of the correlation between $\Delta T$ and $\Delta m_s$ may depend
on the asymmetry energy coefficient and the Coulomb energy coefficient
as suggested in Ref.~\cite{McIntosh13}.
The asymmetry energy and Coulomb energy will depend on these coefficients
as well as the composition of both the QP source and the emitted particle.
By varying these systematically,
it may be possible to extract information leading to constraints on the asymmetry energy.
This would be greatly aided by
a more detailed understanding of
the emission time ordering of LCPs
and its relation to the the ordering
observed in the temperature (Fig.~\ref{fig:CompareLCPTemperature})
and observed in the asymmetry dependence of the temperature (Fig.~\ref{fig:TempDiff_DeltaTvsMs_Lots}).
Extraction of information on the asymmetry energy
would also benefit from an accurate estimate of the density
felt by the probe particles.

\section{Conclusions}

Experimental evidence for the dependence of
nuclear temperatures on the neutron-proton asymmetry
has been presented.
Our previous work \cite{McIntosh13}
has shown a clear decrease in the temperature with the asymmetry
using the momentum quadrupole fluctuation thermometer
with protons as the probe particle.

In this work,
the temperature is extracted with the MQF (kinetic) thermometer
using deuterons, tritons, helions, and alpha particles in addition to protons.
All these probes of the nuclear temperature exhibit
a similar dependence on asymmetry.
The magnitude of the dependence is on the order of 1 MeV
for a change of 0.15 units in asymmetry.
Moreover, this signal is not limited to the MQF method.
We have shown that two Albergo (chemical) thermometers, H/He and Li/He,
provide temperatures that
also clearly exhibit a dependence the asymmetry
on the order of 0.15 MeV for a change of 0.15 units in asymmetry.
For all seven ways of extracting the temperature,
the change in temperature depends linearly on the change in asymmetry.

Additionally,
the ordering of the LCP caloric curves (MQF temperature)
is consistent with an average time-ordering of the LCP emission.
This same ordering is also observed in the
strength of the asymmetry dependence of the temperature.

\begin{acknowledgments}
We thank the staff
of the TAMU Cyclotron Institute for providing the
high quality beams which made this experiment possible.
This work was supported by
the Robert A. Welch Foundation (A-1266),
and the U. S. Department of Energy (DE-FG03-93ER-40773).
\end{acknowledgments}

\bibliography{AsyDepCal_AllLCPs_McIntosh}

\begin{thebibliography}{40}%
\makeatletter
\providecommand \@ifxundefined [1]{%
 \@ifx{#1\undefined}
}%
\providecommand \@ifnum [1]{%
 \ifnum #1\expandafter \@firstoftwo
 \else \expandafter \@secondoftwo
 \fi
}%
\providecommand \@ifx [1]{%
 \ifx #1\expandafter \@firstoftwo
 \else \expandafter \@secondoftwo
 \fi
}%
\providecommand \natexlab [1]{#1}%
\providecommand \enquote  [1]{``#1''}%
\providecommand \bibnamefont  [1]{#1}%
\providecommand \bibfnamefont [1]{#1}%
\providecommand \citenamefont [1]{#1}%
\providecommand \href@noop [0]{\@secondoftwo}%
\providecommand \href [0]{\begingroup \@sanitize@url \@href}%
\providecommand \@href[1]{\@@startlink{#1}\@@href}%
\providecommand \@@href[1]{\endgroup#1\@@endlink}%
\providecommand \@sanitize@url [0]{\catcode `\\12\catcode `\$12\catcode
  `\&12\catcode `\#12\catcode `\^12\catcode `\_12\catcode `\%12\relax}%
\providecommand \@@startlink[1]{}%
\providecommand \@@endlink[0]{}%
\providecommand \url  [0]{\begingroup\@sanitize@url \@url }%
\providecommand \@url [1]{\endgroup\@href {#1}{\urlprefix }}%
\providecommand \urlprefix  [0]{URL }%
\providecommand \Eprint [0]{\href }%
\providecommand \doibase [0]{http://dx.doi.org/}%
\providecommand \selectlanguage [0]{\@gobble}%
\providecommand \bibinfo  [0]{\@secondoftwo}%
\providecommand \bibfield  [0]{\@secondoftwo}%
\providecommand \translation [1]{[#1]}%
\providecommand \BibitemOpen [0]{}%
\providecommand \bibitemStop [0]{}%
\providecommand \bibitemNoStop [0]{.\EOS\space}%
\providecommand \EOS [0]{\spacefactor3000\relax}%
\providecommand \BibitemShut  [1]{\csname bibitem#1\endcsname}%
\let\auto@bib@innerbib\@empty
\bibitem [{\citenamefont {Janka}\ \emph {et~al.}(2007)\citenamefont {Janka}
  \emph {et~al.}}]{Janka07}%
  \BibitemOpen
  \bibfield  {author} {\bibinfo {author} {\bibfnamefont {H.~T.}\ \bibnamefont
  {Janka}} \emph {et~al.},\ }\href@noop {} {\bibfield  {journal} {\bibinfo
  {journal} {Phys. Rep.}\ }\textbf {\bibinfo {volume} {442}},\ \bibinfo {pages}
  {38} (\bibinfo {year} {2007})}\BibitemShut {NoStop}%
\bibitem [{\citenamefont {Danielewicz}\ \emph {et~al.}(2002)\citenamefont
  {Danielewicz}, \citenamefont {Lacey},\ and\ \citenamefont
  {Lynch}}]{Danielewicz02}%
  \BibitemOpen
  \bibfield  {author} {\bibinfo {author} {\bibfnamefont {P.}~\bibnamefont
  {Danielewicz}}, \bibinfo {author} {\bibfnamefont {R.}~\bibnamefont {Lacey}},
  \ and\ \bibinfo {author} {\bibfnamefont {W.~G.}\ \bibnamefont {Lynch}},\
  }\href@noop {} {\bibfield  {journal} {\bibinfo  {journal} {Science}\ }\textbf
  {\bibinfo {volume} {298}},\ \bibinfo {pages} {1592} (\bibinfo {year}
  {2002})}\BibitemShut {NoStop}%
\bibitem [{\citenamefont {Li}\ \emph {et~al.}(2008)\citenamefont {Li},
  \citenamefont {Chen},\ and\ \citenamefont {Ko}}]{Li08}%
  \BibitemOpen
  \bibfield  {author} {\bibinfo {author} {\bibfnamefont {B.~A.}\ \bibnamefont
  {Li}}, \bibinfo {author} {\bibfnamefont {L.~W.}\ \bibnamefont {Chen}}, \ and\
  \bibinfo {author} {\bibfnamefont {C.~M.}\ \bibnamefont {Ko}},\ }\href@noop {}
  {\bibfield  {journal} {\bibinfo  {journal} {Phys. Rep.}\ }\textbf {\bibinfo
  {volume} {464}},\ \bibinfo {pages} {113} (\bibinfo {year}
  {2008})}\BibitemShut {NoStop}%
\bibitem [{\citenamefont {Lattimer}\ and\ \citenamefont
  {Prakash}(2004)}]{Lattimer04}%
  \BibitemOpen
  \bibfield  {author} {\bibinfo {author} {\bibfnamefont {J.~M.}\ \bibnamefont
  {Lattimer}}\ and\ \bibinfo {author} {\bibfnamefont {M.}~\bibnamefont
  {Prakash}},\ }\href@noop {} {\bibfield  {journal} {\bibinfo  {journal}
  {Science}\ }\textbf {\bibinfo {volume} {304}},\ \bibinfo {pages} {536}
  (\bibinfo {year} {2004})}\BibitemShut {NoStop}%
\bibitem [{\citenamefont {Pochodzalla}\ \emph {et~al.}(1995)\citenamefont
  {Pochodzalla} \emph {et~al.}}]{Pochodzalla95}%
  \BibitemOpen
  \bibfield  {author} {\bibinfo {author} {\bibfnamefont {J.}~\bibnamefont
  {Pochodzalla}} \emph {et~al.},\ }\href@noop {} {\bibfield  {journal}
  {\bibinfo  {journal} {Phys. Rev. Lett.}\ }\textbf {\bibinfo {volume} {75}},\
  \bibinfo {pages} {1040} (\bibinfo {year} {1995})}\BibitemShut {NoStop}%
\bibitem [{\citenamefont {Natowitz}\ \emph {et~al.}(2002)\citenamefont
  {Natowitz} \emph {et~al.}}]{Natowitz02}%
  \BibitemOpen
  \bibfield  {author} {\bibinfo {author} {\bibfnamefont {J.~B.}\ \bibnamefont
  {Natowitz}} \emph {et~al.},\ }\href@noop {} {\bibfield  {journal} {\bibinfo
  {journal} {Phys. Rev. C}\ }\textbf {\bibinfo {volume} {65}},\ \bibinfo
  {pages} {034618} (\bibinfo {year} {2002})}\BibitemShut {NoStop}%
\bibitem [{\citenamefont {McIntosh}\ \emph {et~al.}(2013)\citenamefont
  {McIntosh} \emph {et~al.}}]{McIntosh13}%
  \BibitemOpen
  \bibfield  {author} {\bibinfo {author} {\bibfnamefont {A.~B.}\ \bibnamefont
  {McIntosh}} \emph {et~al.},\ }\href {\doibase 10.1016/j.physletb.2012.12.073}
  {\bibfield  {journal} {\bibinfo  {journal} {Phys. Lett. B}\ } (\bibinfo
  {year} {2013}),\ 10.1016/j.physletb.2012.12.073},\ \bibinfo {note} {in
  press}\BibitemShut {NoStop}%
\bibitem [{\citenamefont {Kolomietz}\ \emph {et~al.}(2001)\citenamefont
  {Kolomietz}, \citenamefont {Sanzhur}, \citenamefont {Shlomo},\ and\
  \citenamefont {Firin}}]{Kolomietz01}%
  \BibitemOpen
  \bibfield  {author} {\bibinfo {author} {\bibfnamefont {V.~M.}\ \bibnamefont
  {Kolomietz}}, \bibinfo {author} {\bibfnamefont {A.~I.}\ \bibnamefont
  {Sanzhur}}, \bibinfo {author} {\bibfnamefont {S.}~\bibnamefont {Shlomo}}, \
  and\ \bibinfo {author} {\bibfnamefont {S.~A.}\ \bibnamefont {Firin}},\
  }\href@noop {} {\bibfield  {journal} {\bibinfo  {journal} {Phys. Rev. C}\
  }\textbf {\bibinfo {volume} {64}},\ \bibinfo {pages} {024315} (\bibinfo
  {year} {2001})}\BibitemShut {NoStop}%
\bibitem [{\citenamefont {Hoel}\ \emph {et~al.}(2007)\citenamefont {Hoel},
  \citenamefont {Sobotka},\ and\ \citenamefont {Charity}}]{Hoel07}%
  \BibitemOpen
  \bibfield  {author} {\bibinfo {author} {\bibfnamefont {C.}~\bibnamefont
  {Hoel}}, \bibinfo {author} {\bibfnamefont {L.~G.}\ \bibnamefont {Sobotka}}, \
  and\ \bibinfo {author} {\bibfnamefont {R.~J.}\ \bibnamefont {Charity}},\
  }\href@noop {} {\bibfield  {journal} {\bibinfo  {journal} {Phys. Rev. C}\
  }\textbf {\bibinfo {volume} {75}},\ \bibinfo {pages} {017601} (\bibinfo
  {year} {2007})}\BibitemShut {NoStop}%
\bibitem [{\citenamefont {Besprosvany}\ and\ \citenamefont
  {Levit}(1989)}]{Besprosvany89}%
  \BibitemOpen
  \bibfield  {author} {\bibinfo {author} {\bibfnamefont {J.}~\bibnamefont
  {Besprosvany}}\ and\ \bibinfo {author} {\bibfnamefont {S.}~\bibnamefont
  {Levit}},\ }\href@noop {} {\bibfield  {journal} {\bibinfo  {journal} {Phys.
  Lett. B}\ }\textbf {\bibinfo {volume} {217}},\ \bibinfo {pages} {1} (\bibinfo
  {year} {1989})}\BibitemShut {NoStop}%
\bibitem [{\citenamefont {Ogul}\ and\ \citenamefont {Botvina}(2002)}]{Ogul02}%
  \BibitemOpen
  \bibfield  {author} {\bibinfo {author} {\bibfnamefont {R.}~\bibnamefont
  {Ogul}}\ and\ \bibinfo {author} {\bibfnamefont {A.~S.}\ \bibnamefont
  {Botvina}},\ }\href@noop {} {\bibfield  {journal} {\bibinfo  {journal} {Phys.
  Rev. C}\ }\textbf {\bibinfo {volume} {66}},\ \bibinfo {pages} {051601(R)}
  (\bibinfo {year} {2002})}\BibitemShut {NoStop}%
\bibitem [{\citenamefont {Su}\ and\ \citenamefont {Zhang}(2011)}]{Su11}%
  \BibitemOpen
  \bibfield  {author} {\bibinfo {author} {\bibfnamefont {J.}~\bibnamefont
  {Su}}\ and\ \bibinfo {author} {\bibfnamefont {F.~S.}\ \bibnamefont {Zhang}},\
  }\href@noop {} {\bibfield  {journal} {\bibinfo  {journal} {Phys. Rev. C}\
  }\textbf {\bibinfo {volume} {84}},\ \bibinfo {pages} {037601} (\bibinfo
  {year} {2011})}\BibitemShut {NoStop}%
\bibitem [{\citenamefont {Sfienti}\ \emph {et~al.}(2009)\citenamefont {Sfienti}
  \emph {et~al.}}]{Sfienti09}%
  \BibitemOpen
  \bibfield  {author} {\bibinfo {author} {\bibfnamefont {C.}~\bibnamefont
  {Sfienti}} \emph {et~al.},\ }\href@noop {} {\bibfield  {journal} {\bibinfo
  {journal} {Phys. Rev. Lett.}\ }\textbf {\bibinfo {volume} {102}},\ \bibinfo
  {pages} {152701} (\bibinfo {year} {2009})}\BibitemShut {NoStop}%
\bibitem [{\citenamefont {Colonna}\ and\ \citenamefont
  {Tsang}(2006)}]{Colonna06}%
  \BibitemOpen
  \bibfield  {author} {\bibinfo {author} {\bibfnamefont {M.}~\bibnamefont
  {Colonna}}\ and\ \bibinfo {author} {\bibfnamefont {M.~B.}\ \bibnamefont
  {Tsang}},\ }\href@noop {} {\bibfield  {journal} {\bibinfo  {journal} {Eur.
  Phys. J. A}\ }\textbf {\bibinfo {volume} {30}},\ \bibinfo {pages} {165}
  (\bibinfo {year} {2006})}\BibitemShut {NoStop}%
\bibitem [{\citenamefont {Shetty}\ \emph {et~al.}(2007)\citenamefont {Shetty},
  \citenamefont {Yennello},\ and\ \citenamefont {Souliotis}}]{Shetty07}%
  \BibitemOpen
  \bibfield  {author} {\bibinfo {author} {\bibfnamefont {D.~V.}\ \bibnamefont
  {Shetty}}, \bibinfo {author} {\bibfnamefont {S.~J.}\ \bibnamefont
  {Yennello}}, \ and\ \bibinfo {author} {\bibfnamefont {G.~A.}\ \bibnamefont
  {Souliotis}},\ }\href@noop {} {\bibfield  {journal} {\bibinfo  {journal}
  {Phys. Rev. C}\ }\textbf {\bibinfo {volume} {76}},\ \bibinfo {pages} {024606}
  (\bibinfo {year} {2007})}\BibitemShut {NoStop}%
\bibitem [{\citenamefont {Marini}\ \emph
  {et~al.}(2012{\natexlab{a}})\citenamefont {Marini} \emph
  {et~al.}}]{Marini12}%
  \BibitemOpen
  \bibfield  {author} {\bibinfo {author} {\bibfnamefont {P.}~\bibnamefont
  {Marini}} \emph {et~al.},\ }\href@noop {} {\bibfield  {journal} {\bibinfo
  {journal} {Phys. Rev. C}\ }\textbf {\bibinfo {volume} {85}},\ \bibinfo
  {pages} {034617} (\bibinfo {year} {2012}{\natexlab{a}})}\BibitemShut
  {NoStop}%
\bibitem [{\citenamefont {Trautmann}\ \emph {et~al.}(2008)\citenamefont
  {Trautmann} \emph {et~al.}}]{Trautmann08}%
  \BibitemOpen
  \bibfield  {author} {\bibinfo {author} {\bibfnamefont {W.}~\bibnamefont
  {Trautmann}} \emph {et~al.},\ }\href@noop {} {\bibfield  {journal} {\bibinfo
  {journal} {Int. J. Mod. Phys. E}\ }\textbf {\bibinfo {volume} {17}},\
  \bibinfo {pages} {1838} (\bibinfo {year} {2008})}\BibitemShut {NoStop}%
\bibitem [{\citenamefont {Wuenschel}\ \emph {et~al.}(2010)\citenamefont
  {Wuenschel} \emph {et~al.}}]{Wuenschel10}%
  \BibitemOpen
  \bibfield  {author} {\bibinfo {author} {\bibfnamefont {S.}~\bibnamefont
  {Wuenschel}} \emph {et~al.},\ }\href@noop {} {\bibfield  {journal} {\bibinfo
  {journal} {Nucl. Phys. A}\ }\textbf {\bibinfo {volume} {843}},\ \bibinfo
  {pages} {1} (\bibinfo {year} {2010})}\BibitemShut {NoStop}%
\bibitem [{\citenamefont {Wuenschel}(2009)}]{Wuenschel09_Thesis}%
  \BibitemOpen
  \bibfield  {author} {\bibinfo {author} {\bibfnamefont {S.~K.}\ \bibnamefont
  {Wuenschel}},\ }\href@noop {} {Ph.D. thesis},\ \bibinfo  {school} {Texas A\&M
  Univ.} (\bibinfo {year} {2009})\BibitemShut {NoStop}%
\bibitem [{\citenamefont {Kohley}(2010)}]{Kohley10}%
  \BibitemOpen
  \bibfield  {author} {\bibinfo {author} {\bibfnamefont {Z.}~\bibnamefont
  {Kohley}},\ }\href@noop {} {Ph.D. thesis},\ \bibinfo  {school} {Texas A\&M
  Univ.} (\bibinfo {year} {2010})\BibitemShut {NoStop}%
\bibitem [{\citenamefont {Kohley}\ \emph {et~al.}(2011)\citenamefont {Kohley}
  \emph {et~al.}}]{Kohley11_LCP}%
  \BibitemOpen
  \bibfield  {author} {\bibinfo {author} {\bibfnamefont {Z.}~\bibnamefont
  {Kohley}} \emph {et~al.},\ }\href@noop {} {\bibfield  {journal} {\bibinfo
  {journal} {Phys. Rev. C}\ }\textbf {\bibinfo {volume} {83}},\ \bibinfo
  {pages} {044601} (\bibinfo {year} {2011})}\BibitemShut {NoStop}%
\bibitem [{\citenamefont {Wuenschel}\ \emph
  {et~al.}(2009{\natexlab{a}})\citenamefont {Wuenschel} \emph
  {et~al.}}]{Wuenschel09_NIM}%
  \BibitemOpen
  \bibfield  {author} {\bibinfo {author} {\bibfnamefont {S.}~\bibnamefont
  {Wuenschel}} \emph {et~al.},\ }\href@noop {} {\bibfield  {journal} {\bibinfo
  {journal} {Nucl. Inst. Meth. A}\ }\textbf {\bibinfo {volume} {604}},\
  \bibinfo {pages} {578} (\bibinfo {year} {2009}{\natexlab{a}})}\BibitemShut
  {NoStop}%
\bibitem [{\citenamefont {Wuenschel}\ \emph
  {et~al.}(2009{\natexlab{b}})\citenamefont {Wuenschel} \emph
  {et~al.}}]{Wuenschel09_Isoscaling}%
  \BibitemOpen
  \bibfield  {author} {\bibinfo {author} {\bibfnamefont {S.}~\bibnamefont
  {Wuenschel}} \emph {et~al.},\ }\href@noop {} {\bibfield  {journal} {\bibinfo
  {journal} {Phys. Rev. C}\ }\textbf {\bibinfo {volume} {79}},\ \bibinfo
  {pages} {061602} (\bibinfo {year} {2009}{\natexlab{b}})}\BibitemShut
  {NoStop}%
\bibitem [{\citenamefont {Steckmeyer}\ \emph {et~al.}(2001)\citenamefont
  {Steckmeyer} \emph {et~al.}}]{Steckmeyer01}%
  \BibitemOpen
  \bibfield  {author} {\bibinfo {author} {\bibfnamefont {J.~C.}\ \bibnamefont
  {Steckmeyer}} \emph {et~al.},\ }\href@noop {} {\bibfield  {journal} {\bibinfo
   {journal} {Nucl Phys. A}\ }\textbf {\bibinfo {volume} {686}},\ \bibinfo
  {pages} {537} (\bibinfo {year} {2001})}\BibitemShut {NoStop}%
\bibitem [{\citenamefont {Hinde}\ \emph
  {et~al.}(1989{\natexlab{a}})\citenamefont {Hinde} \emph {et~al.}}]{Hinde89a}%
  \BibitemOpen
  \bibfield  {author} {\bibinfo {author} {\bibfnamefont {D.~J.}\ \bibnamefont
  {Hinde}} \emph {et~al.},\ }\href@noop {} {\bibfield  {journal} {\bibinfo
  {journal} {Phys. Rev. C}\ }\textbf {\bibinfo {volume} {39}},\ \bibinfo
  {pages} {2268} (\bibinfo {year} {1989}{\natexlab{a}})}\BibitemShut {NoStop}%
\bibitem [{\citenamefont {Hinde}\ \emph
  {et~al.}(1989{\natexlab{b}})\citenamefont {Hinde}, \citenamefont {Hilscher},\
  and\ \citenamefont {Rossner}}]{Hinde89b}%
  \BibitemOpen
  \bibfield  {author} {\bibinfo {author} {\bibfnamefont {D.~J.}\ \bibnamefont
  {Hinde}}, \bibinfo {author} {\bibfnamefont {D.}~\bibnamefont {Hilscher}}, \
  and\ \bibinfo {author} {\bibfnamefont {H.}~\bibnamefont {Rossner}},\
  }\href@noop {} {\bibfield  {journal} {\bibinfo  {journal} {Nucl. Phys. A}\
  }\textbf {\bibinfo {volume} {502}},\ \bibinfo {pages} {497} (\bibinfo {year}
  {1989}{\natexlab{b}})}\BibitemShut {NoStop}%
\bibitem [{\citenamefont {Marini}\ \emph
  {et~al.}(2012{\natexlab{b}})\citenamefont {Marini} \emph
  {et~al.}}]{Marini12b}%
  \BibitemOpen
  \bibfield  {author} {\bibinfo {author} {\bibfnamefont {P.}~\bibnamefont
  {Marini}} \emph {et~al.},\ }\href@noop {} {\bibfield  {journal} {\bibinfo
  {journal} {arXiv}\ ,\ \bibinfo {pages} {1211.2134}} (\bibinfo {year}
  {2012}{\natexlab{b}})},\ \bibinfo {note} {accepted for publication in Nucl.
  Inst. Meth. A}\BibitemShut {NoStop}%
\bibitem [{\citenamefont {Wada}\ \emph {et~al.}(2004)\citenamefont {Wada} \emph
  {et~al.}}]{Wada04}%
  \BibitemOpen
  \bibfield  {author} {\bibinfo {author} {\bibfnamefont {R.}~\bibnamefont
  {Wada}} \emph {et~al.},\ }\href@noop {} {\bibfield  {journal} {\bibinfo
  {journal} {Phys. Rev. C}\ }\textbf {\bibinfo {volume} {69}},\ \bibinfo
  {pages} {044610} (\bibinfo {year} {2004})}\BibitemShut {NoStop}%
\bibitem [{\citenamefont {Zheng}\ and\ \citenamefont
  {Bonasera}(2011)}]{Zheng11_PLB}%
  \BibitemOpen
  \bibfield  {author} {\bibinfo {author} {\bibfnamefont {H.}~\bibnamefont
  {Zheng}}\ and\ \bibinfo {author} {\bibfnamefont {A.}~\bibnamefont
  {Bonasera}},\ }\href@noop {} {\bibfield  {journal} {\bibinfo  {journal}
  {Phys. Lett. B}\ }\textbf {\bibinfo {volume} {696}},\ \bibinfo {pages} {178}
  (\bibinfo {year} {2011})}\BibitemShut {NoStop}%
\bibitem [{\citenamefont {Stein}\ \emph {et~al.}(2011)\citenamefont {Stein}
  \emph {et~al.}}]{Stein12x}%
  \BibitemOpen
  \bibfield  {author} {\bibinfo {author} {\bibfnamefont {B.}~\bibnamefont
  {Stein}} \emph {et~al.},\ }\href@noop {} {\bibfield  {journal} {\bibinfo
  {journal} {arXiv}\ ,\ \bibinfo {pages} {1111.2965v1}} (\bibinfo {year}
  {2011})}\BibitemShut {NoStop}%
\bibitem [{\citenamefont {Mabiala}\ \emph {et~al.}(2012)\citenamefont {Mabiala}
  \emph {et~al.}}]{Mabiala12}%
  \BibitemOpen
  \bibfield  {author} {\bibinfo {author} {\bibfnamefont {J.}~\bibnamefont
  {Mabiala}} \emph {et~al.},\ }\href@noop {} {\bibfield  {journal} {\bibinfo
  {journal} {arXiv}\ ,\ \bibinfo {pages} {1208.3480}} (\bibinfo {year}
  {2012})}\BibitemShut {NoStop}%
\bibitem [{\citenamefont {Albergo}\ \emph {et~al.}(1985)\citenamefont
  {Albergo}, \citenamefont {Costa}, \citenamefont {Costanzo},\ and\
  \citenamefont {Rubbino}}]{Albergo85}%
  \BibitemOpen
  \bibfield  {author} {\bibinfo {author} {\bibfnamefont {S.}~\bibnamefont
  {Albergo}}, \bibinfo {author} {\bibfnamefont {S.}~\bibnamefont {Costa}},
  \bibinfo {author} {\bibfnamefont {E.}~\bibnamefont {Costanzo}}, \ and\
  \bibinfo {author} {\bibfnamefont {A.}~\bibnamefont {Rubbino}},\ }\href@noop
  {} {\bibfield  {journal} {\bibinfo  {journal} {Il Nuovo Cimento}\ }\textbf
  {\bibinfo {volume} {89}},\ \bibinfo {pages} {1} (\bibinfo {year}
  {1985})}\BibitemShut {NoStop}%
\bibitem [{\citenamefont {Xi}\ \emph {et~al.}(1999)\citenamefont {Xi},
  \citenamefont {Lynch}, \citenamefont {Tsang}, \citenamefont {Friedman},\ and\
  \citenamefont {Durand}}]{Xi99}%
  \BibitemOpen
  \bibfield  {author} {\bibinfo {author} {\bibfnamefont {H.}~\bibnamefont
  {Xi}}, \bibinfo {author} {\bibfnamefont {W.~G.}\ \bibnamefont {Lynch}},
  \bibinfo {author} {\bibfnamefont {M.~B.}\ \bibnamefont {Tsang}}, \bibinfo
  {author} {\bibfnamefont {W.~A.}\ \bibnamefont {Friedman}}, \ and\ \bibinfo
  {author} {\bibfnamefont {D.}~\bibnamefont {Durand}},\ }\href@noop {}
  {\bibfield  {journal} {\bibinfo  {journal} {Phys. Rev. C}\ }\textbf {\bibinfo
  {volume} {59}},\ \bibinfo {pages} {1567} (\bibinfo {year}
  {1999})}\BibitemShut {NoStop}%
\bibitem [{\citenamefont {Bauer}(1995)}]{Bauer95}%
  \BibitemOpen
  \bibfield  {author} {\bibinfo {author} {\bibfnamefont {W.}~\bibnamefont
  {Bauer}},\ }\href@noop {} {\bibfield  {journal} {\bibinfo  {journal} {Phys.
  Rev. C}\ }\textbf {\bibinfo {volume} {51}},\ \bibinfo {pages} {803} (\bibinfo
  {year} {1995})}\BibitemShut {NoStop}%
\bibitem [{\citenamefont {Kelic}\ \emph {et~al.}(2006)\citenamefont {Kelic},
  \citenamefont {Natowitz},\ and\ \citenamefont {Schmidt}}]{Kelic06}%
  \BibitemOpen
  \bibfield  {author} {\bibinfo {author} {\bibfnamefont {A.}~\bibnamefont
  {Kelic}}, \bibinfo {author} {\bibfnamefont {J.~B.}\ \bibnamefont {Natowitz}},
  \ and\ \bibinfo {author} {\bibfnamefont {K.~H.}\ \bibnamefont {Schmidt}},\
  }\href@noop {} {\bibfield  {journal} {\bibinfo  {journal} {Eur. Phys. J. A}\
  }\textbf {\bibinfo {volume} {30}},\ \bibinfo {pages} {203} (\bibinfo {year}
  {2006})}\BibitemShut {NoStop}%
\bibitem [{\citenamefont {Hudan}\ \emph {et~al.}(2003)\citenamefont {Hudan}
  \emph {et~al.}}]{Hudan03}%
  \BibitemOpen
  \bibfield  {author} {\bibinfo {author} {\bibfnamefont {S.}~\bibnamefont
  {Hudan}} \emph {et~al.},\ }\href@noop {} {\bibfield  {journal} {\bibinfo
  {journal} {arXiv}\ ,\ \bibinfo {pages} {0308031v2}} (\bibinfo {year}
  {2003})}\BibitemShut {NoStop}%
\bibitem [{\citenamefont {Chen}\ \emph {et~al.}(2003)\citenamefont {Chen} \emph
  {et~al.}}]{Chen03}%
  \BibitemOpen
  \bibfield  {author} {\bibinfo {author} {\bibfnamefont {L.~W.}\ \bibnamefont
  {Chen}} \emph {et~al.},\ }\href@noop {} {\bibfield  {journal} {\bibinfo
  {journal} {Nucl. Phys. A}\ }\textbf {\bibinfo {volume} {729}},\ \bibinfo
  {pages} {809} (\bibinfo {year} {2003})}\BibitemShut {NoStop}%
\bibitem [{\citenamefont {Ghetti}\ \emph {et~al.}(2006)\citenamefont {Ghetti}
  \emph {et~al.}}]{Ghetti06}%
  \BibitemOpen
  \bibfield  {author} {\bibinfo {author} {\bibfnamefont {R.}~\bibnamefont
  {Ghetti}} \emph {et~al.},\ }\href@noop {} {\bibfield  {journal} {\bibinfo
  {journal} {Nucl. Phys. A}\ }\textbf {\bibinfo {volume} {765}},\ \bibinfo
  {pages} {307} (\bibinfo {year} {2006})}\BibitemShut {NoStop}%
\bibitem [{\citenamefont {Kohley}\ \emph {et~al.}(2012)\citenamefont {Kohley}
  \emph {et~al.}}]{Kohley12}%
  \BibitemOpen
  \bibfield  {author} {\bibinfo {author} {\bibfnamefont {Z.}~\bibnamefont
  {Kohley}} \emph {et~al.},\ }\href@noop {} {\bibfield  {journal} {\bibinfo
  {journal} {Phys. Rev. C}\ }\textbf {\bibinfo {volume} {86}},\ \bibinfo
  {pages} {044605} (\bibinfo {year} {2012})}\BibitemShut {NoStop}%
\bibitem [{\citenamefont {Borderie}\ \emph {et~al.}(2012)\citenamefont
  {Borderie} \emph {et~al.}}]{Borderie12}%
  \BibitemOpen
  \bibfield  {author} {\bibinfo {author} {\bibfnamefont {A.}~\bibnamefont
  {Borderie}} \emph {et~al.},\ }\href@noop {} {\bibfield  {journal} {\bibinfo
  {journal} {arXiv}\ ,\ \bibinfo {pages} {1207.6085}} (\bibinfo {year}
  {2012})}\BibitemShut {NoStop}%
\end{thebibliography}%

\end{document}